\documentclass[10pt,
prd,
twocolumn,
nofootinbib,
noshowkeys,
noshowpacs,
superscriptaddress,
floatfix
]{revtex4-2}
\usepackage{amsmath,amsfonts,amsthm,amssymb,mathtools}
\usepackage[dvips]{graphics,graphicx}
\usepackage[usenames,dvipsnames]{xcolor}
\definecolor{darkblue}{RGB}{0,0,196}
\definecolor{darkgreen}{RGB}{0,120,0}
\usepackage[colorlinks=true,linktocpage=true,linkcolor=darkblue,citecolor=red,urlcolor=darkblue]{hyperref}
\usepackage{cancel}
\usepackage{bbold}
\usepackage{multirow}
\usepackage{combelow}
\usepackage{longtable}
\usepackage{color}
\usepackage[normalem]{ulem}
\usepackage{bigints}
\usepackage{xparse}
\usepackage{physics}
\usepackage{verbatim}
\usepackage{minibox}
\usepackage{comment}
\usepackage{appendix}
\usepackage{marginnote}
\usepackage{graphicx}
\usepackage{dsfont}
\usepackage{slashed}
\usepackage[nice]{nicefrac}
\usepackage{tikz}
\usepackage{pgfplots}
\pgfplotsset{compat=1.18}
\pgfplotsset{/pgfplots/half and half legend/.style={
        /pgfplots/legend image code/.code={%
            \draw[##1,/tikz/.cd,densely dashed]
            (0cm,0cm) -- (0.35cm,0cm);
            \draw[##1,/tikz/.cd,solid]
            (0.35cm,0cm) -- (0.6cm,0cm);},},              
}
\usepackage{amsmath}
\newcommand{\bk}{\boldsymbol{\kappa}}
\newcommand{\bo}{\boldsymbol{\omega}}
\newcommand{\bn}{\boldsymbol{\nabla}}
\newcommand{\s}{\mathfrak{s}}
\renewcommand{\d}{\mathrm{d}}
\renewcommand{\a}{a}
\renewcommand{\b}{b}

\newcommand{\ft}{\widetilde{f}}
\def\cs{c_\s}
\begin{document}
\preprint{}
\title{Damping of spin waves}

\author{David Wagner}
\email{dwagner@itp.uni-frankfurt.de}
\affiliation{Institut f\"ur Theoretische Physik, 
	Johann Wolfgang Goethe--Universit\"at,
	Max-von-Laue-Str.\ 1, D--60438 Frankfurt am Main, Germany}
\affiliation{Department of Physics, West University of Timi\cb{s}oara, \\
Bd.~Vasile P\^arvan 4, Timi\cb{s}oara 300223, Romania}
\author{Masoud Shokri} 
\email{shokri@itp.uni-frankfurt.de}
\affiliation{Institut f\"ur Theoretische Physik, 
	Johann Wolfgang Goethe--Universit\"at,
	Max-von-Laue-Str.\ 1, D--60438 Frankfurt am Main, Germany}
\author{Dirk H.\ Rischke}
\email{drischke@itp.uni-frankfurt.de}
\affiliation{Institut f\"ur Theoretische Physik, 
	Johann Wolfgang Goethe--Universit\"at,
	Max-von-Laue-Str.\ 1, D--60438 Frankfurt am Main, Germany}
\affiliation{Helmholtz Research Academy Hesse for FAIR, Campus Riedberg,\\
	Max-von-Laue-Str.~12, D-60438 Frankfurt am Main, Germany}
%

%

\newcommand{\bb}[1]{\textcolor{blue}{#1}}
\date{\today} 
%
\begin{abstract}
We show that, in ideal-spin hydrodynamics, the components of the spin tensor follow damped wave equations. The damping rate is related to nonlocal collisions of the particles in the fluid, which enter at first order in $\hbar$ in a semi-classical expansion. 
This rate provides an estimate for the timescale of spin equilibration and is computed by considering a system of spin-1/2 fermions interacting via a quartic self-interaction as well as via (screened) one-gluon exchange. 
It is found that the relaxation times of the components of the spin tensor can become very large compared to the usual dissipative timescales of the system. 
Our results suggest that the spin degrees of freedom in a heavy-ion collision may not be in equilibrium by the time of freeze-out, and thus should be treated dynamically.
\end{abstract}
\maketitle

\section{Introduction}
The well-known Barnett effect~\cite{Barnett:1935} predicts that a rotating fluid consisting of particles with nonvanishing dipole moment becomes polarized along the rotation axis.
A similar effect was conjectured to occur in noncentral heavy-ion collisions, where the large orbital angular momentum gives rise to a nonvanishing vorticity of the hot and dense matter created in the collision, which in turn leads to a nonvanishing polarization of $\Lambda$ hyperons~\cite{Liang:2004ph,Voloshin:2004ha,Betz:2007kg,Becattini:2007sr}. 
This theoretical prediction was subsequently discovered experimentally~\cite{STAR:2017ckg,Adam:2018ivw,ALICE:2019aid}. 
A quantitative explanation of the observed data has initiated many theoretical investigations in recent years~\cite{Florkowski:2017ruc,Florkowski:2017dyn,Hidaka:2017auj,Florkowski:2018myy,Weickgenannt:2019dks,Bhadury:2020puc,Weickgenannt:2020aaf,Shi:2020htn,Speranza:2020ilk,Bhadury:2020cop,Singh:2020rht,Bhadury:2021oat,Peng:2021ago,Sheng:2022ssd,Hu:2021pwh,Hu:2022lpi,Fang:2022ttm,Wang:2022yli,Montenegro:2018bcf,Montenegro:2020paq,Gallegos:2021bzp,Hattori:2019lfp,Fukushima:2020ucl,Li:2020eon,She:2021lhe,Wang:2021ngp,Wang:2021wqq,Hongo:2021ona,Singh:2022ltu,Daher:2022xon,Weickgenannt:2022zxs,Weickgenannt:2022jes,Bhadury:2022ulr,Torrieri:2022ogj,Wagner:2022gza,Weickgenannt:2022qvh,Daher:2022wzf, Xie:2023gbo, Shi:2023sxh, Biswas:2023qsw,Wagner:2023cct,Becattini:2023ouz,Weickgenannt:2023btk,Weickgenannt:2023bss,Weickgenannt:2023nge}.

The theoretical computations of the polarization of particles often neglect the dynamics that cause the alignment of the spin and the vorticity. 
Instead, the spin degrees of freedom are assumed to be in equilibrium, such that the spin tensor of the medium can be expressed through standard fluid-dynamical gradients, such as vorticity and shear \cite{Becattini:2013vja,Fu:2021pok,Becattini:2021iol}. 
In this way, the polarization can be described by ordinary hydrodynamics, without invoking any additional equations of motion.
Such a simplification is justified if the timescale of spin alignment is sufficiently short compared to the other characteristic time scales in the system. 
In general, though, this need not be true, and the evolution of the spin tensor is governed by the conservation equation of the total angular momentum, which provides six additional equations to be solved.
A fluid-dynamical theory that incorporates this conservation law (and possibly additional equations determining the dissipative spin degrees of freedom) is referred to as spin hydrodynamics.

The question of whether the timescale governing the spin-relaxation process is negligibly short is of eminent importance for the correct computation of the polarization: If the timescale is long enough for the dynamics of the spin tensor to play a role in the finally observable polarization of particles, it is mandatory to include its evolution in the hydrodynamic treatment of the system.

Recently, several works have taken on the task of computing the spin-relaxation timescale~\cite{Kapusta:2019sad,Ayala:2019iin,Ayala:2020ndx,Kapusta:2020npk,Hongo:2022izs,Hu:2022xjn,Ayala:2023vgv,Hidaka:2023oze}, with results depending on the microscopic theory employed.
Considering spin-flip interactions of strange quarks in a quark-gluon plasma (QGP), Ref.\ \cite{Kapusta:2019sad} found that the respective relaxation time is very long compared to the lifetime of the system. 
On the other hand, the results of Ref.\ \cite{Hidaka:2023oze} suggest that the time scale of spin relaxation for $\Lambda$ hyperons is on the order of the QGP's lifetime, necessitating a dynamical treatment of the spin degrees of freedom.

References \cite{Weickgenannt:2022zxs,Weickgenannt:2022qvh} computed the timescales on which the \emph{dissipative} components of the spin tensor relax to their so-called Navier-Stokes values, and found them to be comparable to the standard timescales of dissipative hydrodynamics, such as, e.g., the relaxation time of the shear-stress tensor. 
However, in that work, the relaxation times of the \emph{ideal} components of the spin tensor, i.e., the ones whose dynamics are determined by the conservation equation of the total angular momentum, have not been computed. 
The purpose of this paper is to provide an estimate for the latter, using quantum kinetic theory. 

As has been shown in Refs.\ \cite{Ambrus:2022yzz, Hu:2022xjn,Hu:2022mvl,Weickgenannt:2023btk}, the ideal components of the spin tensor follow wave-type equations. 
The damping of the associated spin waves is usually associated with dissipation, but, as has been discussed in Refs.\ \cite{Hu:2022xjn,Hu:2022mvl} in a relaxation-time approximation of the Boltzmann equation, the inclusion of nonlocal collisions induces similar effects. 
In this work, we will explicitly compute such damping through the nonlocal collisions occurring at first order in $\hbar$ in a semi-classical expansion, using a scalar four-fermion interaction as well as (screened) one-gluon exchange and considering the Lorentz-covariant spacetime shifts computed in Ref.\ \cite{Wagner:2022amr}.

In this work, we set $c\equiv k_B\coloneqq 1$, but keep the reduced Planck constant $\hbar$ to indicate the order of quantum effects.
The metric tensor is defined as $g_{\mu\nu}\coloneqq \mathrm{diag}(1,-1,-1,-1)$. 
The scalar product of two four-vectors $a^\mu$, $b^\mu$ is denoted as $a \cdot b \coloneqq a^\mu b_\mu$.
The scalar product of a four-vector $a^\mu$ with the vector $\gamma^\mu$ of Dirac gamma matrices is written as
$\slashed{a} \coloneqq a^\mu \gamma_\mu$.
The antisymmetrization of a rank-two tensor $A$ is defined as $A^{[\mu\nu]}\coloneqq A^{\mu\nu}-A^{\nu\mu}$, while its symmetrization reads $A^{(\mu\nu)}\coloneqq A^{\mu\nu}+A^{\nu\mu}$. 
We denote the comoving derivative of a quantity $X$ as $\dot{X}\coloneqq u \cdot \partial X$.
The four-velocity $u^\mu$ is normalized as $u \cdot u \equiv 1$. The projector orthogonal to $u^\mu$ reads $\Delta^{\mu\nu}\coloneqq g^{\mu\nu}-u^\mu u^\nu$, and a projected vector $a^\mu$ is denoted by $a^{\langle\mu\rangle}\coloneqq \Delta^{\mu\nu}a_\nu$.
Similarly, the traceless symmetric projector of rank four reads $\Delta^{\mu\nu}_{\alpha\beta}\coloneqq \Delta^{(\mu}_\alpha \Delta^{\nu)}_\beta/2 - \Delta^{\mu\nu}\Delta_{\alpha\beta}/3$, and we denote projected tensors by angular brackets, $A^{\langle\mu\nu\rangle}\coloneqq \Delta^{\mu\nu}_{\alpha\beta} A^{\alpha\beta}$.

\section{Ideal-spin hydrodynamics}
The basic equations of spin hydrodynamics in the case of an uncharged fluid are given by the conservation laws for the energy-momentum tensor as well as the total angular-momentum tensor,
\begin{equation}
    \partial_\mu T^{\mu\nu}=0\;,\qquad \partial_\lambda J^{\lambda\mu\nu}=0\;.
    \label{eq:cons_1}
\end{equation}
The total angular-momentum tensor can be decomposed as $J^{\lambda \mu \nu} \coloneqq L^{\lambda \mu \nu} + \hbar S^{\lambda \mu \nu}$, where $L^{\lambda\mu\nu}\coloneqq T^{\lambda[\nu}x^{\mu]}$ is the tensor of orbital angular momentum and $S^{\lambda\mu\nu}$ is the spin tensor. 
Inserting this decomposition, the second conservation law in Eq. \eqref{eq:cons_1} takes the form
\begin{equation}
    \hbar\partial_\lambda S^{\lambda\mu\nu}=T^{[\nu\mu]}\;.
    \label{eq:cons_2}
\end{equation}
In the general case of a dissipative fluid, this system of $4+6=10$ equations is not closed, because the energy-momentum tensor (which does not need to be symmetric) has $16$ components, while the spin tensor has $24$, which follows from the antisymmetry in the last two indices, $S^{\lambda\mu\nu}=-S^{\lambda\nu\mu}$. 
The symmetric part of the energy-momentum tensor can be decomposed as
\begin{equation}
    \frac12 T^{(\mu\nu)}=\varepsilon u^\mu u^\nu - P \Delta^{\mu\nu}+\Pi^{\mu\nu}\;,
    \label{eq:decomp_T}
\end{equation}
where $u^\mu$ is the four-velocity of the fluid, $\varepsilon$ and $P$ denote the energy density and pressure in local equilibrium, and $\Pi^{\mu\nu}$ collects the dissipative degrees of freedom.
In the case of a conventional ideal fluid, the dissipative degrees of freedom vanish, $\Pi^{\mu\nu}=0$. 
Furthermore, the pressure $P$ is not an independent quantity but depends on $\varepsilon$ via an equation of state, $P(\varepsilon)$, such that the ideal energy-momentum tensor is completely characterized by the four fields $\{\varepsilon,\,u^\mu\}$. 
Alternatively, since energy density and pressure in local equilibrium are functions of the temperature $T$, the energy-momentum tensor can equivalently be described by the fields $\{T,u^\mu\}$. 
Thus, in conventional uncharged ideal fluids, the conservation equation for $T^{\mu\nu}$ suffices to describe the fluid evolution.
On the other hand, for a dissipative fluid, one requires additional equations to determine $\Pi^{\mu \nu}$.

In a similar manner, we define an \emph{ideal-spin fluid} as a fluid for which, for a given energy-momentum tensor, Eq.\ (\ref{eq:cons_2}) fully determines the evolution of the spin tensor.
Then, the number of independent components of the spin tensor $S^{\lambda\mu\nu}$ must reduce to six, which can be collected into an antisymmetric second-rank tensor $\Omega^{\mu\nu}$ called the \emph{spin potential}.
The spin potential is a Lagrange multiplier for total angular momentum, just as the inverse four-temperature $\beta^\mu \coloneqq \beta u^\mu$, with $\beta \coloneqq 1/T$, is the Lagrange multiplier for energy-momentum.
The other 18 independent (and purely dissipative) quantities in $S^{\lambda \mu \nu}$ are neglected.
Note that ``ideal'' in our definition refers only to the spin part, since $T^{\mu \nu}$ may still contain dissipative terms, if $\Pi^{\mu \nu} \neq 0$.
The resulting theory of \emph{ideal-spin hydrodynamics} consists of Eq.\ (\ref{eq:cons_2}), the energy-momentum conservation equation, plus additional equations to determine the dissipative quantities in $\Pi^{\mu \nu}$, but neglects all dissipative terms in $S^{\lambda \mu \nu}$.
In global thermodynamic equilibrium, $\Omega^{\mu\nu} =\varpi^{\mu\nu} =\text{const.}$, with $\varpi^{\mu\nu}\coloneqq \frac12\partial^{[\nu}\beta^{\mu]}$ being the thermal vorticity.\footnote{Note that in global equilibrium $\beta_\mu$ is a Killing vector, i.e., $\partial_{(\mu}\beta_{\nu)}=0$.}
In contrast, in ideal-spin hydrodynamics, the spin potential remains an independent variable, which follows an evolution equation.

To leading order, $S^{\lambda\mu\nu}$ is linearly proportional to the spin potential, such that the tensor decomposition of the spin tensor reads
\begin{align}
S^{\lambda\mu\nu}&=A  u^\lambda \Omega^{\mu\nu}+B u^\lambda u_\alpha \Omega^{\alpha[\mu} u^{\nu]} + C u^\lambda \Omega^{\alpha[\mu}\Delta^{\nu]}{}_\alpha\nonumber\\
&\quad+Du_\alpha \Omega^{\alpha[\mu}\Delta^{\nu]\lambda}+E\Delta^\lambda{}_\alpha \Omega^{\alpha[\mu}u^{\nu]}\nonumber\\
&= \left(A-B-C\right) u^\lambda u^{[\mu} \kappa^{\nu]} +E u^{[\mu} \epsilon^{\nu]\lambda\alpha\beta} u_\alpha\omega_{\beta} \nonumber\\
&\quad +\left(A-2C\right)u^\lambda \epsilon^{\mu\nu\alpha\beta} u_\alpha \omega_{\beta} +D \kappa^{[\mu} \Delta^{\nu]\lambda} \;,\label{eq:decomp_S}
\end{align} 
where the quantities $A,\ldots{},E$ are functions of the temperature only and we have used the following decomposition of the spin potential
\begin{equation}
\Omega^{\mu\nu}=u^{[\mu}\kappa^{\nu]}+\epsilon^{\mu\nu\alpha\beta}u_\alpha \omega_{\beta}\;,
\label{eq:decomp_Omega}
\end{equation}
with the ``electric'' and ``magnetic'' components
\begin{equation}
\kappa^\mu \coloneqq -\Omega^{\mu\nu}u_\nu \;,\qquad \omega^\mu \coloneqq \frac12\epsilon^{\mu\nu\alpha\beta}u_\nu \Omega_{\alpha\beta}\;,
\end{equation}
respectively.
Note that the functional forms of the coefficients $A,\ldots{},E$ in Eq.\ \eqref{eq:decomp_S} depend on the underlying microscopic theory.
We will explicitly compute them below in the framework of kinetic theory.
Assuming that the antisymmetric part of the energy-momentum tensor is also linear in $\kappa^\mu$ and $\omega^\mu$, it can be decomposed as (cf.\ Refs.\ \cite{Wang:2021ngp,Daher:2024ixz}, and Appendix \ref{app:EM_Tensor})
\begin{align}
    T^{[\mu\nu]}&=-\hbar^2\Gamma^{(\kappa)} u^{[\mu} \left(\kappa^{\nu]}+ \beta \dot{u}^{\nu ]}\right) \nonumber\\
    &\quad + \hbar^2\Gamma^{(\omega)} \epsilon^{\mu\nu\rho\sigma}u_\rho \left(\omega_\sigma+\beta \Omega_\sigma \right)+\hbar^2 \Pi_A^{\mu\nu}\;,
    \label{eq:T_A}
\end{align}
where $\Omega^\mu\coloneqq \frac{1}{2} \epsilon^{\mu\nu\alpha\beta} u_\nu \nabla_\alpha u_\beta$ is the fluid vorticity vector. 
Furthermore, $\Pi_A^{\mu\nu}=-\Pi_A^{\nu\mu}$ contains dissipative corrections, which, in an ideal-spin fluid, are functions of the standard dissipative degrees of freedom $\Pi^{\mu\nu}$.
Note that, as the system evolves towards global equilibrium, up to a factor $- \beta$ the electric (magnetic) component $\kappa^\mu$ ($\omega^\mu$) of the spin potential approaches the fluid acceleration (the fluid vorticity), while $\Pi_A^{\mu \nu}$ tends to zero.
Then, $T^{[\mu \nu]} \equiv 0$ in global equilibrium. 
Under these conditions, Eq.~(\ref{eq:cons_2}) gives rise to the following constraint on the coefficients $A,\ldots{},E$:
\begin{equation}\label{eq:coeffs_constraint}
   B-C-D-\beta\pdv{E}{\beta} = 0\;.
\end{equation}
The derivation of this relation is shown in Appendix \ref{app:constraint}.
The coefficients $\Gamma^{(\kappa)}$ and $\Gamma^{(\omega)}$ are determined by the underlying microscopic theory, see below.
In Eq.\ \eqref{eq:T_A} we have also assumed that $T^{[\mu \nu]}$ is of order $\hbar^2$, which we will confirm by an explicit calculation in the kinetic-theory framework. 
Up to first order in $\hbar$, the conservation equation for energy and momentum is then simply
\begin{equation}\label{eq:hydro-approx}
 \partial_\mu T^{(\mu\nu)} \approx 0\;.  
\end{equation}
Considering Eq.\ (\ref{eq:decomp_T}), we observe that the spin potential does not enter this equation. 
Thus, the evolution of the fluid fields $\{T, u^\mu\}$, and that of possible dissipative terms in $\Pi^{\mu \nu}$, decouples from the evolution of the spin potential.
The latter is solely determined by Eq.~(\ref{eq:cons_2}), which requires the fluid fields as input.

In the following, we assume the simplest solution to Eq.\ (\ref{eq:hydro-approx}): a fluid with constant temperature at rest, i.e., a static background with $T = \text{const.}$, $u^\mu=(1,\vb{0})^\mu$.
Since the coefficients $A, \ldots , E$ and $\Gamma^{(\kappa)}$, $\Gamma^{(\omega)}$ are functions of temperature only, they also become constant.

\section{Wave equations in a static background}
In a static background, all derivatives of the four-velocity and the temperature vanish, as do the dissipative quantities, $\Pi^{\mu\nu}=\Pi^{\mu\nu}_A=0$.
Then, we insert the divergence of Eq.\ (\ref{eq:decomp_S}) on the left-hand side of Eq.\ \eqref{eq:cons_2} and Eq.\ \eqref{eq:T_A} on the right-hand side.
We contract the resulting equation with $u_\mu$ and $\frac12 \epsilon_{\mu\nu\alpha\beta}u^\beta$ to find the following equations of motion for the components of the spin potential,
\begin{subequations}
\label{eqs:eom_Omega_1}
\begin{align}
(A-B-C)\dot{\kappa}^{\langle\mu\rangle} &= E\epsilon^{\mu\nu\alpha\beta}u_\nu \nabla_\alpha \omega_{\beta}+\hbar \Gamma^{(\kappa)} \kappa^\mu \;,\label{eq:eom_kappa_1}\\
(A-2C)\dot{\omega}^{\langle\mu\rangle} &=D\epsilon^{\mu\nu\alpha\beta} u_\nu \nabla_\alpha \kappa_{\beta} -\hbar \Gamma^{(\omega)} \omega^\mu\;.\label{eq:eom_omega_1}
\end{align}
\end{subequations}
Here we also made use of the fact that $\varpi^{\mu\nu}=0$ in a fluid at rest.
Defining $\kappa^\mu=(0,\bk)^\mu$ and $\omega^\mu=(0,\bo)^\mu$, the equations above can be cast in the form,
\begin{subequations}
\label{eqs:eom_Omega_2}
\begin{align}
    \tau_\kappa \dot{\bk}+\bk &= \mu_\kappa \bn\times \bo\;,\label{eq:eom_kappa_2}\\
    \tau_\omega \dot{\bo}+\bo &= -\mu_\omega \bn\times \bk\;,\label{eq:eom_omega_2}
\end{align}
\end{subequations}
where we defined 
\begin{alignat}{3}
\label{eqs:tau_mu}
    \tau_\kappa&\coloneqq -\frac{A-B-C}{\hbar \Gamma^{(\kappa)}}\;,\quad &&\mu_\kappa\coloneqq -\frac{E}{\hbar \Gamma^{(\kappa)}}\;,\nonumber\\
    \tau_\omega&\coloneqq \frac{A-2C}{\hbar \Gamma^{(\omega)}} \;,\quad &&\mu_\omega\coloneqq -\frac{D}{\hbar \Gamma^{(\omega)}}\;.
\end{alignat}
At this point, it is already apparent that (provided $\tau_\kappa$ and $\tau_\omega$ are positive) both $\bk$ and $\bo$ follow coupled relaxation-type equations, with the characteristic timescales given by $\tau_\kappa$ and $\tau_\omega$.
Taking the divergence of Eqs.\ \eqref{eqs:eom_Omega_2}, we find that the longitudinal components $\bn \cdot \bk$ and $\bn \cdot \bo$ relax to zero on timescales of $\tau_\kappa$ and $\tau_\omega$, respectively. 
Note that this result is consistent with the findings of Ref.\ \cite{Ambrus:2022yzz}, where the relaxation times $\tau_\kappa$ and $\tau_\omega$ diverge, such that the longitudinal components are constant in time.

Let us now consider the dynamics of the transverse degrees of freedom in the case $\bn\cdot \bk= \bn \cdot \bo=0$. 
Then, taking the time derivative of Eqs.\ \eqref{eqs:eom_Omega_2} we obtain 
    \begin{align}\label{eqs:eom_wave}
        \ddot{\mathbf{X}}+ \a \dot{\mathbf{X}} + \b \mathbf{X} - \cs^2 \Delta \mathbf{X} &=0\;,
    \end{align}
where $\mathbf{X}$ is either $\bk$ or $\bo$ and where we defined
\begin{equation}
    \a \coloneqq \frac{\tau_\kappa +\tau_\omega}{\tau_\kappa \tau_\omega} \;,\quad \b\coloneqq \frac{1}{\tau_\kappa \tau_\omega}\;,\quad \cs^2\coloneqq \frac{\mu_\kappa \mu_\omega}{\tau_\kappa \tau_\omega}\;.
\end{equation}
Equations \eqref{eqs:eom_wave} constitute damped wave equations. 
For a symmetric energy-momentum tensor, $\Gamma^{(\kappa)}=\Gamma^{(\omega)}=0$, the coefficients $\a$ and $\b$ vanish, while $\cs^2$ stays finite. 
One thus obtains undamped wave equations as in Ref.\ \cite{Ambrus:2022yzz}.

We now Fourier-Laplace transform the wave equations \eqref{eqs:eom_wave},
\begin{align}
\mathbf{X} (\mathbf{x},t)&=\int_0^\infty \frac{\d s}{2\pi}\int \frac{\d^3 k}{(2\pi)^3} \widetilde{\mathbf{X}}(\mathbf{k},s) e^{-i\mathbf{k}\cdot \mathbf{x}} e^{-st}\;,
\end{align}
and find the dispersion relations
\begin{align}
s_\pm&=\frac{1}{2\tau_\kappa \tau_\omega}\left[\tau_\kappa+\tau_\omega \pm i\sqrt{4\cs^2\tau_\kappa^2 \tau_\omega^2 k^2-(\tau_\kappa-\tau_\omega)^2}\right]\;.
\end{align}
For wavenumbers $k>|\tau_\kappa^{-1}-\tau_\omega^{-1}|/(2\cs)$, $s$ acquires an imaginary part and the spin waves propagate, while for $0\leq k\leq |\tau_\kappa^{-1}-\tau_\omega^{-1}|/(2\cs)$ they only decay. 
Note that in the limit $k\to 0$ we have
\begin{equation} \label{eq:s_kto0}
    s_+ = \frac{1}{\mathrm{max}(\tau_\omega,\tau_\kappa)}\,,\; s_-= \frac{1}{\mathrm{min}(\tau_\omega,\tau_\kappa)}\;,
\end{equation}
indicating that $s_+$ dominates the dynamics of the system at long wavelengths and late times.
In the following, we will explicitly compute $\tau_{\kappa}$ and $\tau_\omega$ using quantum kinetic theory.

\section{Kinetic theory with scalar interaction}
In kinetic theory, the symmetric part of the energy-momentum tensor and the spin tensor are given by \cite{Weickgenannt:2022qvh,Weickgenannt:2022zxs} \footnote{For the sake of brevity we omitted a term that does not contribute to the equations of motion.} 
\begin{subequations}
    \begin{align}
       \frac12  T^{(\mu\nu)}&=\int \mathrm{d}\Gamma k^\mu k^\nu f(x,k,\s)\;,\label{eq:T_kin}\\
        S^{\lambda\mu\nu}&=-\frac{1}{2m}\int \mathrm{d}\Gamma k^\lambda \epsilon^{\mu\nu\alpha\beta} k_\alpha \s_\beta f(x,k,\s)\;.\label{eq:S_kin}
    \end{align}
\end{subequations}
Here, $k^\mu = (k^0, \mathbf{k})^\mu$ is the particle's four-momentum, where $k^0 \equiv \sqrt{\mathbf{k}^2 + m^2}$, with $m$ being the particle mass, and $\mathrm{d}\Gamma\coloneqq \mathrm{d}K \mathrm{d}S$ denotes the integration measure in momentum and spin space, where $\mathrm{d} K \coloneqq \mathrm{d}^3 k/[(2 \pi \hbar)^3 k^0]$, $\mathrm{d} S \coloneqq m/\sqrt{3\pi^2} \mathrm{d}^4 \s\, \delta(k \cdot \s) \delta(\s \cdot \s + 3)$, cf., e.g., Refs.\ \cite{Weickgenannt:2020aaf, Weickgenannt:2021cuo, Wagner:2023cct}.
The quantity $f(x,k,\s)= f_{\text{eq}}(x,k,\s) + \delta f(x,k)$ is the single-particle distribution function, where
\begin{equation}
    f_{\text{eq}} \coloneqq e^{-\beta \cdot k} \left(1  -\frac{\hbar}{4 m} \epsilon^{\mu\nu\alpha\beta} \Omega_{\mu\nu} k_\alpha \s_\beta\right)
    \label{eq:f_eq}
\end{equation}
is the local-equilibrium distribution function for spin-1/2 particles, while $\delta f (x,k)$ contains dissipative corrections and, for an ideal-spin fluid, is independent of $\s$.
In a static background, $\delta f (x,k)$ vanishes and $\beta^\mu$ is constant, while $\Omega_{\mu \nu}$ can be space-time dependent.
Note that, for the sake of simplicity, we neglect quantum statistics.
Inserting the distribution function \eqref{eq:f_eq} into Eq.\ \eqref{eq:T_kin} and using the relation $\int \mathrm{d} S\, \s^\mu = 0$ yields precisely the form \eqref{eq:decomp_T}, with $\varepsilon=I_{20}\,,\; P=I_{21}\,, \;\Pi^{\mu \nu} \equiv 0$, where we introduced the thermodynamic integrals \cite{Denicol:2012cn}
\begin{equation}
    I_{nq}\coloneqq \frac{1}{(2q+1)!!} \int \mathrm{d} \Gamma E_\mathbf{k}^{n-2q} (E_\mathbf{k}^2-m^2)^q e^{-\beta E_{\mathbf{k}}}\;.
\end{equation}
Here, $E_{\mathbf{k}} \coloneqq u \cdot k$ is the energy of a particle with three-momentum $\mathbf{k}$ in the fluid rest frame.
Similarly, inserting the distribution function
\eqref{eq:f_eq} into the spin tensor \eqref{eq:S_kin} and using the relations $\int \mathrm{d} S = 2$ and $\int \mathrm{d} S\, \s^\mu \s^\nu = - 2 (g^{\mu \nu} - k^\mu k^\nu/m^2)$, we obtain Eq.~\eqref{eq:decomp_S}, with the coefficients
\begin{equation}
    A=\frac{\hbar}{4}I_{10}\;,\quad B=\frac{\hbar}{4m^2} I_{30}\;,\quad C=D=E=-\frac{\hbar}{4m^2}I_{31}\;,
\end{equation}
which satisfy the constraint \eqref{eq:coeffs_constraint}.
Upon using these results as well as the relation $m^2 I_{10}=I_{30}-3I_{31}$ in the general expressions \eqref{eqs:tau_mu}, we find
\begin{equation}
    \tau_\kappa = \frac{I_{31}}{2m^2 \Gamma^{(\kappa)}}\;,\quad     \tau_\omega = \frac{I_{30}-I_{31}}{4m^2 \Gamma^{(\omega)}} \;,
    \label{eqs:tau_mu_kin}
\end{equation}
whereas $\mu_\kappa = \tau_\kappa/2$,
$\mu_\omega = \mu_\kappa \Gamma^{(\kappa)}/\Gamma^{(\omega)}$.
We remark at this point that $\cs^2$
is not affected by $T^{[\mu\nu]}$ and agrees with the result of Ref.\ \cite{Ambrus:2022yzz}, if we reduce the latter to classical statistics.

One can show (cf.\ Appendix \ref{app:EM_Tensor}) that, on a static background, the antisymmetric part of the energy-momentum tensor has the form
\begin{align}
T^{[\mu\nu]} &= \frac12 \int [\d\Gamma] \widetilde{\mathcal{W}} \Delta^{[\mu}k^{\nu]}\left(f_{\text{eq},1} f_{\text{eq},2}-f_{\text{eq}} f_{\text{eq}}'\right)\;,\label{eq:T_antisymm_explicit}
\end{align}
where we defined $[\mathrm{d}\Gamma]\coloneqq \mathrm{d}\Gamma_1 \,\mathrm{d} \Gamma_2\, \mathrm{d}\Gamma'\, \mathrm{d}\Gamma$. 
Furthermore, $\widetilde{\mathcal{W}}$ is the transition amplitude for the scattering of two particles from a state with momenta $k_1,k_2$ and spins $\s_1,\s_2$ into a state with momenta $k,k'$ and spins $\s,\s'$.
Finally, $\Delta^\mu$ is the spacetime shift, which quantifies the nonlocality of the collision. 
In Refs.\ \cite{Weickgenannt:2020aaf, Weickgenannt:2021cuo, Wagner:2023cct}, it was found that such a term, which arises as a first-order correction in an $\hbar$-gradient expansion, can describe the mutual conversion of orbital and spin angular momentum. 
The precise form of the transition amplitude and the nonlocality depends on the microscopic interaction of the particles \cite{Wagner:2022amr}. 

Inserting Eq.\ \eqref{eq:f_eq} into Eq.\ (\ref{eq:T_antisymm_explicit}), the antisymmetric part of the energy-momentum tensor becomes
\begin{align}
T^{[\mu\nu]} &= -\frac{\hbar}{8m} \int [\d\Gamma] \widetilde{\mathcal{W}}  \Delta^{[\mu}k^{\nu]} e^{-\beta(E_{\mathbf{k}}+E_{\mathbf{k}'})} \widetilde{\Omega}_{\alpha\beta}\nonumber\\
&\quad \times \left(k_1^\alpha \s_1^\beta +k_2^\alpha \s_2^\beta -k^\alpha \s^\beta -k'^\alpha \s'^\beta\right)\;,\label{eq:T_antisymm_explicit_2}
\end{align}
where we defined $\widetilde{\Omega}^{\mu\nu}\coloneqq  \epsilon^{\mu\nu\alpha\beta}\Omega_{\alpha\beta}$.
Then, we can write it in the form \eqref{eq:T_A}, with the coefficients
\begin{subequations}
\label{eqs:coeffs_expl}
\begin{align}
\Gamma^{(\kappa)}&\coloneqq \frac{1}{12m^2} \int [\d\Gamma] \widetilde{\mathcal{W}}  \frac{m}{\hbar}\Delta^{[\mu}k^{\nu]}e^{-\beta(E_{\mathbf{k}}+E_{\mathbf{k}'})} u_\mu  u^\sigma \epsilon_{\nu\sigma\alpha\beta}  \nonumber\\
&\quad \times\left(k^\alpha \s^\beta +k'^\alpha \s'^\beta -k_1^\alpha \s_1^\beta -k_2^\alpha \s_2^\beta\right) \;,\\
\Gamma^{(\omega)}&\coloneqq \frac{1}{24m^2} \int [\d\Gamma] \widetilde{\mathcal{W}}  \frac{m}{\hbar}\Delta^{[\mu}k^{\nu]} e^{-\beta(E_{\mathbf{k}}+E_{\mathbf{k}'})}\Delta^\rho_\mu \Delta^\sigma_\nu \epsilon_{\rho\sigma\alpha\beta} \nonumber\\
&\quad \times   \left(k^\alpha \s^\beta +k'^\alpha \s'^\beta-k_1^\alpha \s_1^\beta -k_2^\alpha \s_2^\beta\right) \;.
\end{align}
\end{subequations}
We remark that, although both $\Gamma^{(\kappa)}$ and $\Gamma^{(\omega)}$ diverge as $1/m^2$ in the limit of small masses, the coefficients \eqref{eqs:tau_mu_kin} stay finite.
Note that, if the collisions are purely local, i.e., if $\Delta^\mu=0$, the energy-momentum tensor is symmetric. 
Consequently, we then have $\Gamma^{(\kappa)}=\Gamma^{(\omega)}=0$ and the spin waves in Eqs.\ \eqref{eqs:eom_wave} are undamped, which implies that the damping of the spin waves originates solely from the nonlocal part of the collision term. 
The latter drives the relaxation of the spin potential towards the thermal vorticity.

\section{Numerical results}
\begin{figure}
    \centering
    \includegraphics[scale=1]{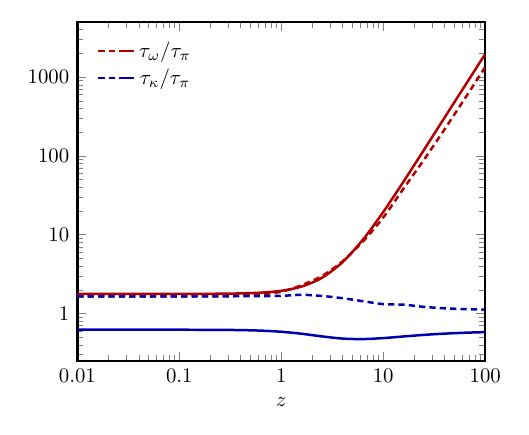}
    \caption{The relaxation times $\tau_\kappa$ and $\tau_\omega$ as  functions of $z=m/T$ in units of the relaxation time of the shear-stress tensor $\tau_{\pi}$. The solid lines denote the result for a scalar four-fermion interaction, while the dashed lines refer to (screened) one-gluon exchange.}
    \label{fig:taus}
\end{figure}
We now compute the relaxation times $\tau_\kappa$ and $\tau_\omega$ for a scalar four-fermion interaction,
$\mathcal{L}_{\text{int}} \coloneqq G (\overline{\psi} \psi)^2$ (for details, see Appendix \ref{app:EM_Tensor}).
In Fig.\ \ref{fig:taus}, we show  $\tau_\kappa$ and $\tau_\omega$ as functions of $z\coloneqq m/T$ in units of the relaxation time $\tau_\pi$ of the shear-stress tensor in the 14-moment approximation \cite{Denicol:2012cn}, computed using the same four-fermion interaction as for $\tau_\kappa$, $\tau_\omega$.
Note that $\tau_\kappa$ is of the same order of magnitude as $\tau_\pi$, as well as the timescales of the dissipative parts of the spin tensor \cite{Weickgenannt:2022qvh, Weickgenannt:2022zxs}.
The ratio $\tau_\kappa/\tau_\pi$ stays approximately constant as a function of $z$. 
On the other hand, the ratio $\tau_\omega/\tau_\pi \sim 2$ at small $z$ and grows like $z^2$, such that $\tau_\omega /\tau_\pi \sim 2 \cdot 10^3$ at $z\approx 100$.
However, the relaxation of \emph{both} $\kappa^\mu$ and $\omega^\mu$, and thus of $\Omega^{\mu \nu}$, is dominated by the \emph{longest} relaxation time, see discussion after Eq.\ \eqref{eq:s_kto0}, which is $\tau_\omega$.
For strange quarks in the deconfined phase of strong-interaction matter, $z \simeq 0.5$, and $\tau_\omega$ is about a factor of four larger than $\tau_\kappa$ and a factor of two larger than $\tau_\pi$. 
In this case, the dynamics of the spin potential occur on similar timescales as the dissipative processes in the system, and one would have to solve the full system of spin-hydrodynamic equations \cite{Weickgenannt:2022qvh,Weickgenannt:2022zxs} to describe the evolution of spin degrees of freedom.
On the other hand, for hyperons in the hadronic phase, we have $z \simeq 10$, and the timescale over which the spin potential $\Omega^{\mu\nu}$ relaxes to its equilibrium value $\varpi^{\mu\nu}$ is considerably larger than the typical dissipative timescales of the system.

We have repeated the calculation of the relaxation times for various interactions (e.g., of vector and axial-vector type), and the increase of $\tau_\omega/\tau_\pi$ with $z$ seems to be a rather robust feature. 
For comparison, in Fig.~\ref{fig:taus} we also show the results for an interaction mediated by one-gluon exchange (and screened by a thermal gluon mass), which is detailed in Appendix \ref{app:EM_Tensor}. 
Note that, in the limit of small $z$, $\tau_\kappa$ and $\tau_\omega$ assume the same value, which is a property of any kind of vector interaction.
On the other hand, for large $z$, $\tau_\omega \gg \tau_\kappa$, as for the scalar four-point interaction.

\section{Conclusion}
The main results of this work are the following three statements: First, in ideal-spin hydrodynamics, the components of the spin tensor in a static fluid background follow wave equations, the solutions of which are \textit{damped} if the energy-momentum tensor has a nonvanishing antisymmetric part. 
Second, if a kinetic description of the fluid is applicable, the damping rate of the spin waves can be expressed through integrals over the nonlocal parts of the collision term. 
Third, an estimate using various types of  interactions shows that the relaxation time for the spin potential towards its equilibrium value varies from timescales of the order of the other dissipative processes (for $z \lesssim 1$) to timescales which are orders of magnitude larger (for $z \gg 1$).

The implication of the third finding for heavy-ion phenomenology is the following: 
Approximating the spin potential with its Navier-Stokes value, i.e., the thermal vorticity, is only justified in the high-temperature or small-mass regime.
In a high-energy heavy-ion collision, the spin potential equilibrates to the thermal vorticity in a hot QGP with light particles.
This explains why the results of Refs.\ \cite{Becattini:2013vja,Fu:2021pok,Becattini:2021iol} are consistent with the data.

However, in order to obtain a complete description of the spin degrees of freedom even in high-energy collisions, one has to treat the spin potential dynamically, at least as long as the masses of the particles carrying the pertinent spin degrees of freedom are of the order of the temperature of the system.
From a computational point of view, this means solving six additional hydrodynamic equations (for $\kappa^\mu$ and $\omega^\mu$) in the ideal case, and several more in the dissipative case, as laid out in Refs.\ \cite{Weickgenannt:2022qvh, Weickgenannt:2022zxs}. 
This becomes even more relevant at lower collision energies, where neglecting the dynamics of the spin potential is certainly a poor approximation.

Our results agree qualitatively with those of previous works that computed relaxation times due to collisions that flip the spin of the particles.
This agreement is to be expected, since this type of collisions corresponds to the nonlocal parts of the collision term in our setup. 
Furthermore, compared to the results of Ref.\ \cite{Hu:2022xjn}, which also considered nonlocal collisions, the separation of our relaxation time scales is larger. We think this difference can be attributed to the form of the collision integral and the use of the Lorentz-covariant nonlocal contributions derived in Ref.\ \cite{Wagner:2022amr}.
We also point out that there is no contradiction to the findings of Refs.\ \cite{Weickgenannt:2022qvh, Weickgenannt:2022zxs}, which obtained rather short relaxation times for the dissipative components of the spin tensor. 
The reason for the difference is that these quantities relax to their Navier-Stokes values through local collisions, which include all types of processes (including those where no spin is flipped) and constitute the dominant contribution to the collision term.

\section*{Acknowledgements}
The authors are supported by the Deutsche Forschungsgemeinschaft (DFG, German Research Foundation) through the CRC-TR 211 ``Strong-interaction matter under extreme conditions'' -- project number 315477589 -- TRR 211, and by the State of Hesse within the Research Cluster ELEMENTS (Project ID 500/10.006). D.W.~acknowledges support by the Ministry of Research, Innovation and Digitization, CNCS - UEFISCDI, project number PN-III-P1-1.1-TE-2021-1707, within PNCDI III.

\appendix 

\section{Derivation of Eq.~(8)}
\label{app:constraint}
Here we derive the constraint on the functions $A,\ldots,E$ given by Eq.~(8) that follows from the fact that the spin tensor should be conserved in global equilibrium. 
Acting with $\partial_\lambda$ on the spin tensor in global equilibrium (i.e., where $\Omega^{\mu\nu}=\varpi^{\mu\nu}$) yields
\begin{align}
    \partial_\lambda S^{\lambda\mu\nu}_{\mathrm{eq}}&=\dot{u}_\alpha\varpi^{\alpha[\mu}u^{\nu]}\left(B-C-E-\beta \frac{\partial E}{\partial \beta}\right) \nonumber\\
    &\quad +u_\alpha \varpi^{\alpha[\mu}\dot{u}^{\nu]} \left(B-C-D-\beta \frac{\partial D}{\partial \beta}\right)\nonumber\\
    &\quad+\varpi_{\alpha}{}^{[\mu}\omega^{\nu]\alpha} (D-E)\;,
\end{align}
where we defined the vorticity $\omega^{\mu\nu}\coloneqq \frac12 \nabla^{[\mu}u^{\nu]}$ and used that in global equilibrium $\nabla^{\mu}u^\nu=\omega^{\mu\nu}$. 
Employing that in global equilibrium
\begin{equation}
    \varpi^{\mu\nu}=-\beta u^{[\mu}\dot{u}^{\nu]} - \beta \omega^{\mu\nu} \;,
\end{equation}
we find that
\begin{equation}
    \varpi_{\alpha}{}^{[\mu}\omega^{\nu]\alpha}=\beta \dot{u}_\alpha u^{[\mu} \omega^{\nu]\alpha}= -  \dot{u}_\alpha u^{[\mu} \varpi^{\nu]\alpha}= -\dot{u}_\alpha \varpi^{\alpha[\mu} u^{\nu]}\;,
\end{equation}
whereas $u_\alpha \varpi^{\alpha[\mu}\dot{u}^{\nu]}=0$. Then, the divergence of the spin tensor reads
\begin{equation}
    \partial_\lambda S^{\lambda\mu\nu}_{\mathrm{eq}}=\dot{u}_\alpha\varpi^{\alpha[\mu}u^{\nu]}\left(B-C-D-\beta \frac{\partial E}{\partial \beta}\right) \;,
\end{equation}
leading to Eq.~(8) in the main text.

\section{The antisymmetric part of the energy-momentum tensor}
\label{app:EM_Tensor}
In this section, we derive the antisymmetric part of the energy-momentum tensor that was used in the main text.
Starting from interacting Dirac fields and employing the Wigner-function formalism, one can show that, to second order in $\hbar$, the conservation equation of the total angular momentum reads \cite{Weickgenannt:2022zxs}
\begin{widetext}
\begin{align}
\hbar \partial_\lambda S^{\lambda\mu\nu} &\equiv
\frac{\hbar}{2} \partial_\lambda \int \d \Gamma k^\lambda \Sigma_\s^{\mu\nu} f(x,k,\s)\nonumber\\
&=\frac{\hbar}{4} \int [\mathrm{d\Gamma}] \widetilde{\mathcal{W}} \Sigma_\s^{\mu\nu}\left[f(x+\Delta_1-\Delta,k_1,\s_1) f(x+\Delta_2-\Delta,k_2,\s_2) \ft(x+\Delta'-\Delta,k',\s') \ft(x,k,\s)\right.\nonumber\\
    &\hspace{2.6cm}\left.- \ft(x+\Delta_1-\Delta,k_1,\s_1)\ft(x+\Delta_2-\Delta,k_2,\s_2) f(x+\Delta'-\Delta,k',\s') f(x,k,\s)\right]\nonumber\\
    &\equiv T^{[\nu\mu]}\;,\label{eq:div_S_T_1}
\end{align}
where we defined $\Sigma_\s^{\mu\nu}\coloneqq -\frac{1}{m}\epsilon^{\mu\nu\alpha\beta} k_\alpha \s_\beta$, $\widetilde{f}\coloneqq 1-f$ (for Fermi-Dirac statistics), and we have employed the Boltzmann equation \cite{Wagner:2022amr}.
Moreover, we have made use of the ``weak-equivalence principle'' of Ref.\ \cite{Weickgenannt:2021cuo} to simplify the collision term.
Furthermore,
\begin{subequations}
\begin{equation} \label{eq:tildeW}
    \widetilde{\mathcal{W}}\coloneqq (2\pi\hbar)^4 \delta^{(4)}(k+k'-k_1-k_2) m^4M^{\alpha_1 \alpha_2 \beta_1 \beta_2}M^{\gamma_1\gamma_2 \delta_1 \delta_2}h_{\beta_1\gamma_1}(k_1,\s_1)h_{\beta_2 \gamma_2}(k_2,\s_2)   h_{\delta_2 \alpha_2}(k',\s')h_{\delta_1\alpha_1}(k,\s) 
\end{equation}
is the transition amplitude \cite{Wagner:2022amr} and
\begin{align}
\Delta_{1}^\mu &\coloneqq-\frac{i\hbar m^3}{4\widetilde{\mathcal{W}}} (2\pi\hbar)^4 \delta^{(4)}(k+k'-k_1-k_2) M^{\alpha_1 \alpha_2 \beta_1 \beta_2}M^{\gamma_1\gamma_2 \delta_1 \delta_2} h_{\beta_2 \gamma_2}(k_2,\s_2)   h_{\delta_2 \alpha_2}(k',\s') h_{\delta_1 \alpha_1}(k,\s) \left[h(k_1,\s_1),\gamma^\mu\right]_{\beta_1 \gamma_1}\;,\label{eq:def_Delta_1_Dirac}\\
\Delta_{2}^\mu &\coloneqq-\frac{i\hbar m^3}{4\widetilde{\mathcal{W}}} (2\pi\hbar)^4 \delta^{(4)}(k+k'-k_1-k_2)  M^{\alpha_1 \alpha_2 \beta_1 \beta_2}M^{\gamma_1\gamma_2 \delta_1 \delta_2} h_{\beta_1\gamma_1}(k_1,\s_1)   h_{\delta_2 \alpha_2}(k',\s') h_{\delta_1 \alpha_1}(k,\s) \left[h(k_2,\s_2),\gamma^\mu\right]_{\beta_2 \gamma_2}\;,\label{eq:def_Delta_2_Dirac}\\
\Delta'^\mu &\coloneqq-\frac{i\hbar m^3}{4\widetilde{\mathcal{W}}} (2\pi\hbar)^4 \delta^{(4)}(k+k'-k_1-k_2) M^{\alpha_1 \alpha_2 \beta_1 \beta_2}M^{\gamma_1\gamma_2 \delta_1 \delta_2} h_{\beta_1\gamma_1}(k_1,\s_1) h_{\beta_2 \gamma_2}(k_2,\s_2)    h_{\delta_1 \alpha_1}(k,\s) \left[h(k',\s'),\gamma^\mu\right]_{\delta_2 \alpha_2}\;,\label{eq:def_Delta_prime_Dirac}\\
\Delta^\mu &\coloneqq-\frac{i\hbar m^3}{4\widetilde{\mathcal{W}}}  (2\pi\hbar)^4 \delta^{(4)}(k+k'-k_1-k_2) M^{\alpha_1 \alpha_2 \beta_1 \beta_2}M^{\gamma_1\gamma_2 \delta_1 \delta_2}h_{\beta_1\gamma_1}(k_1,\s_1)h_{\beta_2 \gamma_2}(k_2,\s_2)   h_{\delta_2 \alpha_2}(k',\s')  \left[h(k,\s),\gamma^\mu\right]_{\delta_1 \alpha_1}\label{eq:def_Delta_Dirac}
\end{align}
\end{subequations}
are the spacetime shifts, with
\begin{equation}
    h_{\alpha\beta}(k,\s)\coloneqq \frac{1}{4m}\left[(\mathds{1}+\gamma_5 \slashed{\s})(\slashed{k}+m)\right]_{\alpha\beta}\;.
\end{equation}
The fourth-rank tensors $M_{\alpha\alpha'\alpha_1\alpha_2}$ describe the vertices of the theory.
In the case of a scalar four-point interaction, $\mathcal{L}_{\text{int}} \coloneqq G (\overline{\psi} \psi)^2$ with coupling strength $G$, they are given by
\begin{equation} \label{eq:vertex}
M_{\alpha\alpha'\alpha_1\alpha_2}= \frac{2G}{\hbar} \left(\delta_{\alpha \alpha_1}\delta_{\alpha'\alpha_2}-\delta_{\alpha \alpha_2}\delta_{\alpha'\alpha_1}\right)\;,
\end{equation}
leading to the following expressions for the transition rate and the spacetime shift:
\begin{subequations} \label{eq:tildeW-Delta_2}
\begin{align}
    \widetilde{\mathcal{W}}&=(2\pi\hbar)^4 \delta^{(4)}(k+k'-k_1-k_2) \frac{8m^4 G^2}{\hbar^2}\mathrm{Re}\left[\mathrm{Tr}\left(hh_2\right)\mathrm{Tr}\left(h_1h'\right)-\mathrm{Tr}\left(hh_1h'h_2\right)\right] 
    \;,\\
    \Delta^\mu_1&= (2\pi\hbar)^4 \delta^{(4)}(k+k'-k_1-k_2)\frac{\hbar}{m}\frac{4 m^4 G^2}{\hbar^2 \widetilde{\mathcal{W}}}\mathrm{Im}\left[\mathrm{Tr}\left(h h_2\right)\mathrm{Tr}\left(h_1 \gamma^\mu h'\right)-\mathrm{Tr}\left(h h_1 \gamma^\mu h'h_2\right)\right]\;,\\
    \Delta^\mu_2&= (2\pi\hbar)^4 \delta^{(4)}(k+k'-k_1-k_2)\frac{\hbar}{m}\frac{4 m^4 G^2}{\hbar^2 \widetilde{\mathcal{W}}}\mathrm{Im}\left[\mathrm{Tr}\left(h h_2\gamma^\mu\right)\mathrm{Tr}\left(h_1h'\right)-\mathrm{Tr}\left(h h_1h'h_2\gamma^\mu\right)\right]\;,\\
    \Delta'^\mu&= (2\pi\hbar)^4 \delta^{(4)}(k+k'-k_1-k_2)\frac{\hbar}{m}\frac{4 m^4 G^2}{\hbar^2 \widetilde{\mathcal{W}}}\mathrm{Im}\left[\mathrm{Tr}\left(h h_2\right)\mathrm{Tr}\left(h_1h'\gamma^\mu\right)-\mathrm{Tr}\left(h h_1h'\gamma^\mu h_2\right)\right]\;,\\
    \Delta^\mu&= (2\pi\hbar)^4 \delta^{(4)}(k+k'-k_1-k_2)\frac{\hbar}{m}\frac{4 m^4 G^2}{\hbar^2 \widetilde{\mathcal{W}}}\mathrm{Im}\left[\mathrm{Tr}\left(h\gamma^\mu h_2\right)\mathrm{Tr}\left(h_1h'\right)-\mathrm{Tr}\left(h\gamma^\mu h_1h'h_2\right)\right]\;.
\end{align}
\end{subequations}
\end{widetext}
Here, the traces are taken over Dirac space, and we abbreviated $h_i\coloneqq h(k_i,\s_i)$, $i = 1,2$,
and similarly for $h$, $h'$.
In order to obtain Eqs.\ \eqref{eq:tildeW-Delta_2}, we have used the fact that $\widetilde{\mathcal{W}}$ and $\Delta^\mu$ are symmetric under the exchange $(k_1,\s_1) \leftrightarrow (k_2,\s_2)$ under the integral in Eq.\ \eqref{eq:div_S_T_1}.
Moreover, we have employed that $h^\dagger=\gamma^0 h \gamma^0$ as well as the fact that, for the vertices \eqref{eq:vertex}, the identity
\begin{equation}
    \gamma^0_{\alpha\beta} \gamma^0_{\alpha'\beta'} M^*_{\beta\beta'\beta_1\beta_2}\gamma^0_{\beta_1\alpha_1}\gamma^0_{\beta_2\alpha_2} = M_{\alpha_1\alpha_2\alpha\alpha'}\label{eq:id_M}
\end{equation}
holds.
In the case of QCD, the interaction Lagrangian between the quarks and gluons reads $\mathcal{L}_{\text{int}}\coloneqq g \overline{\psi}\gamma^\mu \mathcal{A}_\mu \psi$, with the strong coupling constant $g$ and the gluon field $\mathcal{A}^\mu$. 
When considering the lowest-order binary scattering of quarks mediated by thermal gluons, the vertices of the theory depend on momentum and read
\begin{align}
    M_{\alpha\alpha'\alpha_1\alpha_2}&= \frac{2g}{\hbar} \bigg[ \gamma^\mu_{\alpha\alpha_1}\frac{g_{\mu\nu}}{\left(k-k_1\right)^2-m_\text{th}^2}\gamma^\nu_{\alpha'\alpha_2}\nonumber\\
    &\qquad -\gamma^\mu_{\alpha\alpha_2}\frac{g_{\mu\nu}}{\left(k-k_2\right)^2-m_\text{th}^2}\gamma^\nu_{\alpha'\alpha_1}\bigg]\;,\label{eq:M_pQCD}
\end{align}
where $m_\text{th}\coloneqq \sqrt{2N_c+N_f}gT/(3\sqrt{2})$ is the thermal mass of the gluons \cite{Braaten:1990it}. For the computation, we take $N_c=N_f=3$.

We now turn to a different way to express $T^{[\nu\mu]}$ (cf.\ Eq.\ (43a) of Ref.\ \cite{Weickgenannt:2022jes}),
\begin{equation}
    T^{[\nu\mu]} = \frac{\hbar}{m} \int \frac{\d^4 k}{(2\pi\hbar)^4} \mathcal{D}_\mathcal{V}^{[\mu}k^{\nu]}\;,\label{eq:div_S_T_2}
\end{equation}
where $\mathcal{D}_\mathcal{V}^\mu = \Re\langle:\Tr\gamma^\mu C:\rangle$, with the matrix $C$ being the collision term in Dirac space.

Equations \eqref{eq:div_S_T_1} and \eqref{eq:div_S_T_2} provide us with two equivalent descriptions of the antisymmetric part of the energy-momentum tensor; in the following, we will use both of them.
From Eq.~\eqref{eq:div_S_T_1} it is clear that $T^{[\mu\nu]}$ contains both local and nonlocal parts, with the latter being proportional to gradients of the involved distribution functions. 
This suggests that the collision term $\mathcal{D}_\mathcal{V}^\mu$ can also be split into local and nonlocal parts (to be specified in Subsec.\ \ref{subsec:nonlocal_Tmunu}).
Therefore, we can write the local part of $T^{[\mu\nu]}$ as
\begin{align}
    T^{[\mu\nu]}_{\mathrm{local}} &= -\frac{\hbar}{4} \int [\mathrm{d\Gamma}] \widetilde{\mathcal{W}}\Sigma_\s^{\mu\nu} (f_1 f_2 \ft' \ft- \ft_1\ft_2 f' f) \nonumber\\
    &= -\frac{\hbar}{m} \int \frac{\d^4 k}{(2\pi\hbar)^4} \mathcal{D}_{\mathcal{V},\mathrm{local}}^{[\mu}k^{\nu]}\;,\label{eq:Tmunu_local_general}
\end{align}
where we abbreviated $f_1\coloneqq f(x,k_1,\s_1)$ and similarly for $f_2$, $f'$, and $f$.
Note that the second line in this equation results from Eq.\ \eqref{eq:div_S_T_2}.
On the other hand, using Eqs.\ \eqref{eq:div_S_T_1} and \eqref{eq:div_S_T_2}, the nonlocal part is given by
\begin{align}
    T^{[\mu\nu]}_{\mathrm{nonlocal}} &= -\frac{\hbar}{4} \int [\mathrm{d\Gamma}] \widetilde{\mathcal{W}}\Sigma_\s^{\mu\nu}  \nonumber\\
    &\times\bigg\{\left(\Delta_1^\lambda-\Delta^\lambda\right)\left[  (\partial_\lambda f_1) f_2   \ft' \ft -(\partial_\lambda\ft_1)\ft_2 f' f\right] \nonumber\\
    &\quad+\left(\Delta_2^\lambda-\Delta^\lambda\right)\left[f_1 (\partial_\lambda f_2)   \ft' \ft-\ft_1 (\partial_\lambda\ft_2) f' f  \right] \nonumber\\
    &\quad +\left(\Delta'^\lambda-\Delta^\lambda\right)\left[f_1 f_2  (\partial_\lambda \ft') \ft-\ft_1 \ft_2 (\partial_\lambda f') f \right]\bigg\}\nonumber\\
    &= -\frac{\hbar}{m} \int \frac{\d^4 k}{(2\pi\hbar)^4} \mathcal{D}_{\mathcal{V},\mathrm{nonlocal}}^{[\mu}k^{\nu]}\;.\label{eq:Tmunu_nonlocal_general}
\end{align}
In the next subsections, we will evaluate the local and nonlocal terms separately.

\subsection{Local part}
To evaluate the local term of the antisymmetric part of the energy-momentum tensor, we may use both lines of Eq.~\eqref{eq:Tmunu_local_general}, since their equivalence is enforced by the conservation of total angular momentum.
It turns out to be advantageous to explicitly compute the collision term $\mathcal{D}_\mathcal{V}^\mu$, which (to first order in $\hbar$) reads \cite{Wagner:2022amr}
\begin{align}
    \mathcal{D}_\mathcal{V}^\mu &\coloneqq \frac12 \mathrm{Im} \mathrm{Tr} \left[\gamma^\mu \left(\Sigma^> G^< -\Sigma^< G^>\right)\right]\nonumber\\
    &\quad-\frac{\hbar}{4} \mathrm{Re}\mathrm{Tr}\left[\gamma^\mu\left(\{\Sigma^>,G^<\}_{\mathrm{PB}}-\{\Sigma^<,G^>\}_{\mathrm{PB}}\right)\right]\;.
\end{align}
Here, $\Sigma^{\gtrless}$ are the greater and lesser self-energies, and $G^\gtrless$ are the greater and lesser propagators, respectively. 
Furthermore, we defined the Poisson bracket of two functions $f(x,k)$ and $g(x,k)$, 
\begin{equation}
\left\{f(x,k),g(x,k)\right\}_{\mathrm{PB}}\coloneqq (\partial_\mu f)(\partial_k^\mu g)- (\partial_k^\mu f)(\partial_\mu g)\;. 
\end{equation}\\[0.1cm]
As shown in Ref.\ \cite{Wagner:2022amr}, we can expand
the propagators and self-energies in terms of ``quasiclassical'' and ``gradient'' contributions,
\begin{equation}
    G^{\gtrless}=G^\gtrless_{\mathrm{qc}}+G^\gtrless_\nabla\;,\qquad \Sigma^{\gtrless}=\Sigma^\gtrless_{\mathrm{qc}}+\Sigma^\gtrless_\nabla\;.
\end{equation}
Since both the gradient and Poisson-bracket contributions involve derivatives of the distribution functions, the required local part of $\mathcal{D}_\mathcal{V}^\mu$ consists solely of the quasiclassical contributions,
\begin{equation}    \mathcal{D}_{\mathcal{V},\mathrm{local}}^\mu \coloneqq \frac12 \mathrm{Im} \mathrm{Tr} \left[\gamma^\mu \left(\Sigma^>_{\mathrm{qc}} G^<_{\mathrm{qc}} -\Sigma^<_{\mathrm{qc}} G^>_{\mathrm{qc}}\right)\right]\;.\label{eq:D_qc}
\end{equation}
The quasiclassical propagators can be expressed in terms of the distribution functions $f$ and $\ft$ as
\begin{subequations}
\label{eqs:G_qc_Dirac_extended}
\begin{align}
G^<_{\text{qc}}(x,k)&=-4m\pi\hbar \delta(k^2-m^2) \int \d S(k) h(k,\s) f(x,k,\s)\;, \\
G^>_{\text{qc}}(x,k)&= 4m\pi\hbar \delta(k^2-m^2) \int \d S(k) h(k,\s) \widetilde{f}(x,k,\s) \;,
\end{align}
\end{subequations}
where we defined the lesser propagator with a different sign as compared to Ref.\ \cite{Wagner:2022amr}.
The quasiclassical contributions to the self-energies are
\begin{widetext}
\begin{subequations}
\label{eqs:Sigma_qc_extended_Dirac}
\begin{align}
\Sigma^{<}_{\text{qc},\alpha\beta}(x,k)&= \frac{m^3}{2}\int \d \Gamma_1\, \d \Gamma_2\, \d \Gamma' (2\pi\hbar)^4 \delta^{(4)}(k+k'-k_1-k_2)M_{\alpha\alpha'\alpha_1\alpha_2}M_{\beta_1\beta_2\beta\beta'}\nonumber\\
&\quad\times h_{\alpha_1\beta_1}(k_1,\s_1)h_{\alpha_2\beta_2}(k_2,\s_2)h_{\beta'\alpha'}(k',\s') f(x,k_1,\s_1)f(x,k_2,\s_2) \widetilde{f}(x,k',\s')
\end{align}
and
\begin{align}
\Sigma^{>}_{\text{qc},\alpha\beta}(x,k)&= -\frac{m^3}{2}\int \d \Gamma_1\, \d \Gamma_2\, \d \Gamma' (2\pi\hbar)^4 \delta^{(4)}(k+k'-k_1-k_2)M_{\alpha\alpha'\alpha_1\alpha_2}M_{\beta_1\beta_2\beta\beta'}\nonumber\\
&\quad\times h_{\alpha_1\beta_1}(k_1,\s_1)h_{\alpha_2\beta_2}(k_2,\s_2)h_{\beta'\alpha'}(k',\s') \widetilde{f}(x,k_1,\s_1)\widetilde{f}(x,k_2,\s_2) f(x,k',\s') \;,
\end{align}
\end{subequations}
respectively.
Here, the indices $\alpha,\,\beta,\ldots$ are Dirac indices, and the Einstein summation convention is assumed.
Inserting Eqs.\ \eqref{eqs:G_qc_Dirac_extended} and \eqref{eqs:Sigma_qc_extended_Dirac} into Eq. \eqref{eq:D_qc}, we find
\begin{align}\label{eq:dv-local}
    \mathcal{D}_{\mathcal{V},\mathrm{local}}^\mu &= -4m\pi \hbar \delta(k^2-m^2) \int \d \Gamma_1 \,\d \Gamma_2 \, \d \Gamma' \, \d S(k) (2\pi\hbar)^4 \delta^{(4)}(k+k'-k_1-k_2)\, \mathcal{U}^\mu \left(f_1 f_2 \ft' \ft - \ft_1 \ft_2 f' f\right)\;,
\end{align}
where we defined
\begin{align}
\mathcal{U}^\mu &\coloneqq \frac{m^3}{4} \Im \left[M_{\alpha \alpha'\alpha_1\alpha_2} M_{\beta_1\beta_2 \beta \beta'} h_{\alpha_1\beta_1}(k_1,\s_1)h_{\alpha_2\beta_2}(k_2,\s_2) h_{\beta'\alpha'}(k',\s') h_{\beta \delta}(k,\s) \gamma^\mu_{\delta\alpha} \right]\;.
\end{align}
\end{widetext}
Comparing this expression to Eq.\  \eqref{eq:def_Delta_Dirac} and employing the identity \eqref{eq:id_M}, we find
\begin{equation}\label{eq:mcu}
    (2\pi\hbar)^4\delta^{(4)}(k+k'-k_1-k_2)\, \mathcal{U}^\mu = \frac{1}{2\hbar} \widetilde{\mathcal{W}} \Delta^\mu\;.
\end{equation}
Upon inserting Eq.\ \eqref{eq:mcu} into Eq.\ \eqref{eq:dv-local} and subsequently into Eq.\ \eqref{eq:Tmunu_local_general}, we arrive at the result
\begin{equation}
    T^{[\mu\nu]}_{\mathrm{local}} = \frac12 \int [\mathrm{d\Gamma}] \widetilde{\mathcal{W}}\Delta^{[\mu}k^{\nu]} (f_1 f_2 \ft' \ft- \ft_1\ft_2 f' f) \;.\label{eq:Tmunu_local_full}
\end{equation}

Before proceeding, let us note that the result \eqref{eq:Tmunu_local_full} gives a precise meaning to the total angular momentum as a collisional invariant. 
Subtracting the first from the second line of Eq.\ \eqref{eq:Tmunu_local_general}, we find
\begin{equation}
    \frac12 \int [\mathrm{d\Gamma}] \widetilde{\mathcal{W}}\left(\Delta^{[\mu}k^{\nu]}+\frac{\hbar}{2} \Sigma^{\mu\nu}_\s\right) (f_1 f_2 \ft' \ft- \ft_1\ft_2 f' f)=0\;,
\end{equation}
which allows us to identify the collisional invariant as $J^{\mu\nu}\coloneqq \Delta^{[\mu}k^{\nu]}+\frac{\hbar}{2} \Sigma^{\mu\nu}_\s$. 
Indeed, this intuitive form has already been assumed in Refs.\ \cite{Weickgenannt:2021cuo, Weickgenannt:2022qvh}, and here we confirmed it rigorously.

In general, the distribution function can be written as $f=f_{\mathrm{eq}}+\delta f$, where 
\begin{equation}
    f_{\mathrm{eq}}\coloneqq f_0\left(1 +\ft_0\frac{\hbar}{4} \Sigma_\s^{\mu\nu}\Omega_{\mu\nu}\right)+ \mathcal{O}(\hbar^2)
\end{equation}
is the local-equilibrium distribution function, with $f_0$ being the Fermi-Dirac distribution.
Then, the local part of $T^{[\mu\nu]}$ takes the form
\begin{equation}
    T^{[\mu\nu]}_{\mathrm{local}}\equiv T^{[\mu\nu]}_{\mathrm{local},0}+\delta T^{[\mu\nu]}_{\mathrm{local}}\;,
\end{equation}
where $\delta T^{[\mu\nu]}_{\mathrm{local}}$ collects the dissipative contributions and is of no further interest for the following.
Inserting the local-equilibrium distribution function and neglecting terms of second order in $\hbar$, we obtain
\begin{align}
    T^{[\mu\nu]}_{\mathrm{local},0}&= \frac{\hbar}{8}\int [\mathrm{d}\Gamma]  \widetilde{\mathcal{W}}    \ft_{0,1}\ft_{0,2} f'_0 f_0 \Delta^{[\mu} k^{\nu]}\nonumber\\
    &\quad\times\Omega_{\alpha\beta}\left(\Sigma_{\s_1}^{\alpha\beta}+\Sigma_{\s_2}^{\alpha\beta}-\Sigma_{\s'}^{\alpha\beta}-\Sigma_{\s}^{\alpha\beta}\right)\;.\label{eq:Tmunu_qc_local_eq}
\end{align}\\[-0.1cm]

\subsection{Nonlocal part}
\label{subsec:nonlocal_Tmunu}
To compute the nonlocal part of $T^{[\mu\nu]}$, it is simpler to employ the first and second lines of Eq.\ \eqref{eq:Tmunu_nonlocal_general} rather than the third one. 
First, note that we can separate this expression as
\begin{equation}
   T^{[\mu\nu]}_{\mathrm{nonlocal}}=T^{[\mu\nu]}_{\mathrm{nonlocal},1}+T^{[\mu\nu]}_{\mathrm{nonlocal},2} \;,
\end{equation}
where we defined
\begin{widetext}
\begin{subequations}
\label{eqs:Tmunu_nonlocal_1_2}
\begin{align}
T^{[\mu\nu]}_{\mathrm{nonlocal},1} &\coloneqq \frac{\hbar}{4} \int [\mathrm{d\Gamma}] \widetilde{\mathcal{W}}\Sigma_\s^{\mu\nu}\Delta\cdot \partial   \left(f_1 f_2 \ft' \ft- \ft_1\ft_2 f' f\right)\label{eq:Tmunu_nonlocal_1}\;,\\
    T^{[\mu\nu]}_{\mathrm{nonlocal},2} &\coloneqq -\frac{\hbar}{4} \int [\mathrm{d\Gamma}] \widetilde{\mathcal{W}}\Sigma_\s^{\mu\nu}\bigg\{\Delta_1^\lambda\left[  (\partial_\lambda f_1) f_2   \ft' \ft -(\partial_\lambda\ft_1)\ft_2 f' f\right]+\Delta_2^\lambda\left[f_1 (\partial_\lambda f_2)   \ft' \ft-\ft_1 (\partial_\lambda\ft_2) f' f  \right]  \nonumber\\
    &\quad+\Delta'^\lambda\left[f_1 f_2  (\partial_\lambda \ft') \ft-\ft_1 \ft_2 (\partial_\lambda f') f \right]+\Delta^\lambda\left[  f_1 f_2   \ft' (\partial_\lambda \ft) -\ft_1\ft_2 f' (\partial_\lambda f)\right]\bigg\}\;.\label{eq:Tmunu_nonlocal_2}
\end{align}
\end{subequations}
\end{widetext}
We will evaluate these expressions further by neglecting their dissipative contributions as well as terms of higher order in $\hbar$, i.e., we set $f=f_0$, and likewise for $f_1$, $f_2$, and $f'$.
This immediately implies that $T^{[\mu\nu]}_{\mathrm{nonlocal},1}=0$.
Now, we write
\begin{equation}\label{eq:partial_f}
    \partial_\lambda f_0= -f_0 \ft_0 k^\rho \partial_\lambda \beta_\rho=f_0 \ft_0 k^\rho \left(\varpi_{\lambda\rho}-\xi_{\lambda\rho}\right)\;,
\end{equation}
where we defined the thermal vorticity and the thermal shear respectively as 
\begin{equation}
    \varpi^{\mu\nu}\coloneqq -\frac12 \partial^{[\mu}\beta^{\nu]}\;,\qquad \xi^{\mu\nu}\coloneqq \frac12 \partial^{(\mu}\beta^{\nu)}\;.
\end{equation}
Upon inserting Eq.\ \eqref{eq:partial_f} into Eq.\ \eqref{eq:Tmunu_nonlocal_2} we obtain
\begin{widetext}
\begin{equation}
     T^{[\mu\nu]}_{\mathrm{nonlocal},2} = -\frac{\hbar}{4} \int [\mathrm{d\Gamma}] \widetilde{\mathcal{W}}\Sigma_\s^{\mu\nu}\ft_{0,1}\ft_{0,2} f'_0 f_0   \left(\Delta_1^\alpha k_1^\beta+\Delta_2^\alpha k_2^\beta-\Delta'^\alpha k'^\beta-\Delta^\alpha k^\beta\right)\left(\varpi_{\alpha\beta}-\xi_{\alpha\beta}\right)\;.
    \label{eq:Tmunu_nonlocal_2_eq}
\end{equation}
Renaming integration variables and using that $\ft_{0,1} \ft_{0,2} f'_0 f_0= \ft_0 \ft'_0 f_{0,1} f_{0,2}$, we can rewrite this as
\begin{equation}     T^{[\mu\nu]}_{\mathrm{nonlocal},2} =- \frac{\hbar}{4} \int [\mathrm{d\Gamma}] \widetilde{\mathcal{W}}\ft_{0,1}\ft_{0,2} f'_0 f_0   \Delta^\alpha k^\beta\left(\Sigma_{\s_1}^{\mu\nu}+\Sigma_{\s_2}^{\mu\nu}-\Sigma_{\s'}^{\mu\nu}-\Sigma_{\s}^{\mu\nu}\right)\left(\varpi_{\alpha\beta}-\xi_{\alpha\beta}\right)\;.
\end{equation}
Using the fact that the thermal vorticity is antisymmetric and the thermal shear is symmetric, the most general form of this integral is
\begin{equation}
    T^{[\mu\nu]}_{\mathrm{nonlocal},2} = X^{\mu\nu\alpha\beta} \varpi_{\alpha\beta}-Y^{\mu\nu\alpha\beta} \xi_{\alpha\beta}\;,
\end{equation}
where $X^{\mu\nu\alpha\beta}$ ($Y^{\mu\nu\alpha\beta}$) is antisymmetric (symmetric) in $\alpha$ and $\beta$.
Explicitly, these rank-four tensors read
\begin{subequations}
\begin{align}
    X^{\mu\nu\alpha\beta}&\coloneqq -\frac{\hbar}{8} \int [\mathrm{d\Gamma}] \widetilde{\mathcal{W}} \ft_{0,1}\ft_{0,2} f'_0 f_0 \Delta^{[\alpha}k^{\beta]}\left(\Sigma_{\s_1}^{\mu\nu}+\Sigma_{\s_2}^{\mu\nu}-\Sigma_{\s'}^{\mu\nu}-\Sigma_{\s}^{\mu\nu}\right)\;,\\
    Y^{\mu\nu\alpha\beta}&\coloneqq -\frac{\hbar}{8} \int [\mathrm{d\Gamma}] \widetilde{\mathcal{W}} \ft_{0,1}\ft_{0,2} f'_0 f_0 \Delta^{(\alpha}k^{\beta)}\left(\Sigma_{\s_1}^{\mu\nu}+\Sigma_{\s_2}^{\mu\nu}-\Sigma_{\s'}^{\mu\nu}-\Sigma_{\s}^{\mu\nu}\right)\;.
\end{align}
\end{subequations}
\end{widetext}
Since $X^{\mu \nu \alpha \beta}$ and $Y^{\mu \nu \alpha \beta}$ depend only on equilibrium quantities, their respective tensor decompositions read
\begin{subequations}
\begin{align}
    X^{\mu\nu\alpha\beta}&\coloneqq X_1 \, u^{[\alpha}\Delta^{\beta][\mu}u^{\nu]}+ X_2\, g^{[\beta}_\rho \Delta^{\alpha][\mu}\Delta^{\nu]\rho}\;,\\
    Y^{\mu\nu\alpha\beta}&\coloneqq Y_1 \, u^{(\alpha}\Delta^{\beta)[\mu}u^{\nu]}\;,
\end{align}
\end{subequations}
with scalar functions $X_1,\, X_2,\, Y_1$, which can be obtained from the tensors $X^{\mu \nu \alpha \beta}$ and $Y^{\mu \nu \alpha \beta}$ by suitable projections,
\begin{subequations}
    \begin{align}
        X_1 & \coloneqq \frac{1}{3} u_\mu u_\beta \Delta_{\nu \alpha}\,
        X^{\mu \nu \alpha \beta}\;, \quad X_2 \coloneqq \frac{1}{12} \Delta_{\nu \beta} \Delta_{\mu \alpha} \, X^{\mu \nu \alpha \beta}\;, \\
        Y_1 & \coloneqq - \frac{1}{3} u_\mu u_\beta \Delta_{\nu \alpha}\,
        Y^{\mu \nu \alpha \beta}\;.
    \end{align}
\end{subequations}
From these expressions, it becomes clear that $X^{\mu\nu\alpha\beta}=X^{\alpha\beta\mu\nu}$, whereas there is no such relation for $Y^{\mu\nu\alpha\beta}$. 
Therefore, we can write
\begin{equation}
    T^{[\mu\nu]}_{\mathrm{nonlocal},2} = X^{\alpha\beta\mu\nu} \varpi_{\alpha\beta}-Y^{\mu\nu\alpha\beta} \xi_{\alpha\beta}\;,
\end{equation}
or explicitly
\begin{align}
    T^{[\mu\nu]}_{\mathrm{nonlocal},2} &= -\frac{\hbar}{8} \int [\mathrm{d\Gamma}] \widetilde{\mathcal{W}} \ft_{0,1}\ft_{0,2} f'_0 f_0 \Delta^{[\mu}k^{\nu]}\varpi_{\alpha\beta}\nonumber\\
    &\qquad \times\left(\Sigma_{\s_1}^{\alpha\beta}+\Sigma_{\s_2}^{\alpha\beta}-\Sigma_{\s'}^{\alpha\beta}-\Sigma_{\s}^{\alpha\beta}\right)\nonumber\\
    &\quad +  \frac{\hbar}{8} \int [\mathrm{d\Gamma}] \widetilde{\mathcal{W}} \ft_{0,1}\ft_{0,2} f'_0 f_0 \Delta^{(\alpha}k^{\beta)}\xi_{\alpha\beta}\nonumber\\
    &\qquad \times\left(\Sigma_{\s_1}^{\mu\nu}+\Sigma_{\s_2}^{\mu\nu}-\Sigma_{\s'}^{\mu\nu}-\Sigma_{\s}^{\mu\nu}\right)\;.\label{eq:T_munu_nonlocal_eq}
\end{align}

\subsection{Complete expression}
Combining Eqs.\ \eqref{eq:Tmunu_qc_local_eq} and \eqref{eq:T_munu_nonlocal_eq}, we find for the antisymmetric part of the energy-momentum tensor in local equilibrium
\begin{align}
    T^{[\mu\nu]}_0 &= \frac{\hbar}{8} \left(\Omega_{\alpha\beta}-\varpi_{\alpha\beta}\right)\int [\mathrm{d\Gamma}] \widetilde{\mathcal{W}} \ft_{0,1}\ft_{0,2} f'_0 f_0 \Delta^{[\mu}k^{\nu]}\nonumber\\
    &\qquad\times\left(\Sigma_{\s_1}^{\alpha\beta}+\Sigma_{\s_2}^{\alpha\beta}-\Sigma_{\s'}^{\alpha\beta}-\Sigma_{\s}^{\alpha\beta}\right)\nonumber\\
    &\quad +  \frac{\hbar}{8}\xi_{\alpha\beta} \int [\mathrm{d\Gamma}] \widetilde{\mathcal{W}} \ft_{0,1}\ft_{0,2} f'_0 f_0 \Delta^{(\alpha}k^{\beta)}\nonumber\\
    &\qquad\times\left(\Sigma_{\s_1}^{\mu\nu}+\Sigma_{\s_2}^{\mu\nu}-\Sigma_{\s'}^{\mu\nu}-\Sigma_{\s}^{\mu\nu}\right)\;.
\end{align}
As expected, this only vanishes in global equilibrium, where $\Omega^{\mu\nu}=\varpi^{\mu\nu}$ and $\xi^{\mu\nu}=0$.
Given the tensor structures at our disposal, the most general decomposition is
\begin{align}
T^{[\mu\nu]}_0&=\hbar^2\Gamma^{(\kappa)} u^{[\mu} \left(\Omega^{\nu]\alpha}-\varpi^{\nu]\alpha}\right)u_\alpha\nonumber\\
&\quad+ \hbar^2\Gamma^{(\omega)} \left(\Omega^{\langle\mu\rangle\langle\nu\rangle}-\varpi^{\langle\mu\rangle\langle\nu\rangle}\right)+\hbar^2\Gamma^{(a)} u^{[\mu} \xi^{\nu]\alpha}u_\alpha\;,
\end{align}
where we defined the coefficients
\begin{widetext}
\begin{subequations}
\begin{align}
    \Gamma^{(\kappa)}&\coloneqq - 2\frac{X_1}{\hbar^2} =\frac{1}{12\hbar} u_\mu u_\beta \Delta_{\nu\alpha} \int [\mathrm{d\Gamma}] \widetilde{\mathcal{W}} \ft_{1,0}\ft_{2,0} f'_0 f_0 \Delta^{[\mu}k^{\nu]}\left(\Sigma_{\s_1}^{\alpha\beta}+\Sigma_{\s_2}^{\alpha\beta}-\Sigma_{\s'}^{\alpha\beta}-\Sigma_{\s}^{\alpha\beta}\right)\;,\\
    \Gamma^{(\omega)}&\coloneqq  - 4\frac{X_2}{\hbar^2} =\frac{1}{24\hbar} \Delta_{\nu\beta} \Delta_{\mu\alpha} \int [\mathrm{d\Gamma}] \widetilde{\mathcal{W}} \ft_{1,0}\ft_{2,0} f'_0 f_0 \Delta^{[\mu}k^{\nu]}\left(\Sigma_{\s_1}^{\alpha\beta}+\Sigma_{\s_2}^{\alpha\beta}-\Sigma_{\s'}^{\alpha\beta}-\Sigma_{\s}^{\alpha\beta}\right)\;,\\
   \Gamma^{(a)}&\coloneqq  2\frac{Y_1}{\hbar^2} = \frac{1}{12\hbar}u_\mu u_\beta \Delta_{\nu\alpha} \int [\mathrm{d\Gamma}] \widetilde{\mathcal{W}} \ft_{1,0}\ft_{2,0} f'_0 f_0 \Delta^{(\alpha}k^{\beta)}\left(\Sigma_{\s_1}^{\mu\nu}+\Sigma_{\s_2}^{\mu\nu}-\Sigma_{\s'}^{\mu\nu}-\Sigma_{\s}^{\mu\nu}\right)\;.
\end{align}
\end{subequations}
\end{widetext}
However, note that
\begin{align}
    \xi^{\langle\nu\rangle\alpha}u_\alpha &= \frac12\left[\nabla^\nu (\beta u^\alpha) +\Delta^{\nu}_\lambda\partial^\alpha (\beta u^\lambda)\right]u_\alpha\nonumber\\
    &=\frac{1}{2}\left(\nabla^\nu\beta +\beta\dot{u}^\nu\right)\;,
\end{align}
which, together with the relation (for uncharged fluids)
$
    \nabla^\nu \beta=-\beta\nabla^\nu P/ ({\varepsilon+P})
$
and the equation of motion (neglecting dissipative terms)
\begin{equation}
    \dot{u}^\nu=\frac{\nabla^\nu P}{\varepsilon+P}
\end{equation}
yields
\begin{equation}
    \xi^{\langle\nu\rangle\alpha}u_\alpha = 0\;,
\end{equation}
such that we can drop these terms as they are of the same order as dissipative ones. Evaluating also the projections of the spin potential and the thermal vorticity, we finally obtain
\begin{align}
    T^{[\mu\nu]}_0&=-\hbar^2\Gamma^{(\kappa)} u^{[\mu} \left(\kappa^{\nu]}+\beta\dot{u}^{\nu]}\right) \nonumber\\
    &\quad+ \hbar^2\Gamma^{(\omega)} \epsilon^{\mu\nu\alpha\beta}u_\alpha\left(\omega_\beta+\beta\Omega_\beta\right)\;,
\end{align}
where $\Omega^\mu\coloneqq  \frac12 \epsilon^{\mu\nu\alpha\beta} u_\nu \nabla_{\alpha}u_{\beta}$ is the fluid vorticity vector.

\bibliography{bib_spinwaves}
\end{document}